\documentclass[reprint, amsmath,amssymb,aps, prb,superscriptaddress]{revtex4-1}
\usepackage{amsfonts,amssymb}
\usepackage{graphicx}
\usepackage{hyperref}    
\usepackage{bm}          
\usepackage{natbib}
\usepackage{soul} 
\usepackage{color}
\usepackage{longtable}
\usepackage{ textcomp }

\begin{document}

\title{Accumulation,  inversion, and depletion layers in SrTiO$_3$} 

\author{K.V. Reich}
\email{Reich@mail.ioffe.ru}
\affiliation{Fine Theoretical Physics Institute, University of Minnesota, Minneapolis, MN 55455, USA}
\affiliation{Ioffe Institute, St Petersburg, 194021, Russia}
\author{M. Schecter}
\affiliation{Fine Theoretical Physics Institute, University of Minnesota, Minneapolis, MN 55455, USA}
\author{B.I. Shklovskii}
\affiliation{Fine Theoretical Physics Institute, University of Minnesota, Minneapolis, MN 55455, USA}

\begin{abstract}
We study potential and electron density depth profiles in accumulation, inversion and depletion layers in crystals with a large
and nonlinear dielectric response such as $\mathrm{SrTiO_3}$. We describe the lattice dielectric response using the Landau-Ginzburg free energy expansion. In accumulation and inversion layers we arrive at new nonlinear dependencies of the width $d$ of the electron gas on an applied electric field $D_0$. Particularly important is the predicted electron density profile of accumulation layers (including the $\mathrm{LaAlO_3/SrTiO_3}$ interface) $n(x) \propto (x+d)^{-12/7}$, where $d \propto D_0^{-7/5}$. We compare this profile with available data and find satisfactory agreement. For a depletion layer we find an unconventional nonlinear dependence of the capacitance on voltage. We also evaluate the role of spatial dispersion in the dielectric response by adding a gradient term to the Landau-Ginzburg free energy.
\end{abstract}

\pacs{73.20.-r,71.10.Ca,73.30.+y,68.47.Gh}

\date{\today}

\maketitle

\section{Introduction}
\label{sec:introduction}
In recent years, there has been growing interest in the investigation of $\mathrm{ABO_3}$ perovskite crystals, which are important for numerous technological applications and show intriguing magnetic, superconducting, and multiferroic  properties \cite{Oxides_rev}.  Special attention \cite{Stemmer_STO,Zubko_oxides} is paid to heterostructures involving $\mathrm{SrTiO_3}$ which is a  semiconductor  with a band gap of $\simeq \mathrm{3.2~eV}$  \cite{Optical_absorbtion_STO} and a large dielectric constant $\kappa$ ranging from $2 \cdot 10^4$ at liquid helium temperatures to $350$ at room temperature.  As with conventional semiconductors,  $\mathrm{SrTiO_3}$ can be used as a building block for different types of devices, with reasonably large mobility \cite{Ohtomo_2004,Hwang_mobility}.

Many devices are based on the accumulation layer of electrons  near a heterojunction interface involving $\mathrm{SrTiO_3}$. For example, one can use modulation doping in the structure $\mathrm{SrTiO_3}$/$\mathrm{SrZrO_3}$ to introduce electrons in the conduction band of $\mathrm{SrTiO_3}$ from $\mathrm{La}$ donors within the wider-band-gap material $\mathrm{SrZrO_3}$ \cite{modulation_doped}.  Inside bulk  $\mathrm{SrTiO_3}$  $\delta$-doping can be used  to introduce two accumulation layers of electrons  \cite{delta_doped_stemmer,delta_doped_STO_Hwang,delta_doped_STO_Stemmer}. One can  accumulate an electron gas  using a field-effect \cite{10_percent,Hwang_gate,Stemmer_concentration_interface} instead of dopants. In Refs. \onlinecite{induced_superconductivity,Gallagher_2014} the authors accumulated up to $10^{14} ~\mathrm{cm}^{-2}$ electrons on the surface of $\mathrm{SrTiO_3}$  using ionic liquid gating.  Finally, there is enormous interest~\cite{Ohtomo_2004,Hwang_mobility,Stemmer_STO} in $\mathrm{LaAlO_3/SrTiO_3}$  heterojunctions where electrons are accumulated by the electric field resulting from the ``polar catastrophe''  \cite{polar-catastrophe_2006}.  It is natural to  think that the depth profiles of the potential and electron density inside $\mathrm{SrTiO_3}$  have a universal origin in all these devices.

Another type of device based on $n$-doped $\mathrm{SrTiO_3}$ is the Schottky diode. Due to the built-in Schottky barrier the region near the metal-semiconductor interface in doped $\mathrm{SrTiO_3}$ is depleted. The large and nonlinear  dielectric constant results in unconventional capacitance-voltage characteristics. Schottky diodes with different metals and bulk $\mathrm{SrTiO_3}$ dopants have been studied:  $\mathrm{Au / Nb:SrTiO_3}$  \cite{Hwang_Schottky}, $\mathrm{Ba_{1-x}K_xBiO_3 / Nb:SrTiO_3}$ \cite{JJAP.36.L390}, $\mathrm{SrRuO_{3} / Nb:SrTiO_3}$ \cite{Schottky_Hwang}, and $\mathrm{Au/SrTiO_{3-x}}$ \cite{Stemmer_Shottky}.

All of the devices cited above are based on  accumulation and depletion layers. We do not know of any attempts to create a hole inversion layer in $n$-type $\mathrm{SrTiO_3}$ or an electron inversion layer in $p$-type $\mathrm{SrTiO_3}$ but they are likely to be of interest as well.

Interface properties determine characteristics of  all these devices. Not surprisingly, the potential and electron density depth profiles in such devices have attracted attention from the experimental  \cite{Hwang_Xray,Hwang_PL,LAO_STO_Berreman,Stemmer_GdTO} and theoretical points of view \cite{MacDonald_theory,induced_superconductivity, abinitio_STO,abinitio_STO_2,distribution_LAO_STO,superconductivity_LAO_STO}. For example, experimental data show that electrons are distributed in a layer of width $\simeq \mathrm{5-10~nm}$  near the   $\mathrm{LaAlO_3 / SrTiO_3}$ interface. Theoretical works that attempt to explain such behavior are based on microscopic numerical calculations.

The goal of this paper is to create a simple, mostly phenomenological and analytical  approach for describing the potential and the electron density depth profiles  in $\mathrm{SrTiO_3}$. To account for the nonlinear dielectric response in $\mathrm{SrTiO_3}$ we use the Landau-Ginzburg free energy expansion \cite{Ginzburg_ferroelectrics, landau_stat1}. Electrons are almost everywhere described in the Thomas-Fermi approximation \cite{Thomas_Fermi}.  Although we mostly concentrate on $\mathrm{SrTiO_3}$, the developed approach is applicable to $\mathrm{KTaO_3}$ \cite{Rowley_2014} and $\mathrm{CaTiO_3}$ \cite{CaTiO3_quantum} as well. 

Our main result is a new form for  the potential and electron density depth profiles   in accumulation, inversion and depletion layers due to the nonlinear dielectric response. In particular, for an accumulation layer in $\mathrm{SrTiO_3}$, we find an electron concentration $n(x)$ that depends on the distance from the surface $x$ as $n(x) \propto (x+d)^{-12/7}$, where the width $d$ decreases with the  external electric field $D_0$ as $d \propto D_0^{-7/5}$. These relations seem to agree with  experimental data \cite{LAO_STO_Berreman,Hwang_PL}.

The remainder of this paper is organized as follows. In Sec.~\ref{sec:model} we define the model based on the Landau-Ginzburg theory for calculating the lattice dielectric response  and describe the parameters of $\mathrm{SrTiO_3}$.   In Sec.~\ref{sec:accumulation} we use the Thomas-Fermi approach for calculating the self-consistent electric field to describe properties of electron accumulation layers. In the Sec.~\ref{sec:experiment} we apply our theory to the consideration of interfaces between $\mathrm{SrTiO_3}$ and polar dielectrics. In particular, we pay attention to the case of an accumulation layer on the interface of $\mathrm{LaAlO_3/SrTiO_3}$ and compare our theory  with experimental data. In Sec. \ref{sec:quantum} we calculate the quantum capacitance of the accumulation layer in  $\mathrm{SrTiO_3}$. In Sec.~\ref{sec:inversion} we use a one sub-band approximation, in which electrons do not affect the electric field, to calculate properties of inversion layers. In Sec.~\ref{sec:depletion} we consider a depletion layer in $\mathrm{SrTiO_3}$,  calculate the capacitance-voltage characteristics of a Schottky barrier for such systems, and compare our results with  experimental data. In Sec.~\ref{sec:dispersion} we show that our  results are not modified by the presence of spatial dispersion  in the dielectric response. Sec. ~\ref{sec:conclusion} provides a summary and conclusion.

\section{The model}
\label{sec:model}

Bulk $\mathrm{SrTiO_3}$ typically is an $n$-type semiconductor with a concentration of donors  $N>10^{17} ~ \mathrm{cm^{-3}}$. Let us discuss the position of Fermi energy $\varepsilon_F$ in such crystals. The electron spectrum near the bottom of the conduction band is complicated~\cite{Mazin_band_structure}, and in order to make the problem of an accumulation layer tractable analytically we assume that it is isotropic and non-degenerate with the effective mass $m^* \simeq 1.5~ m$, \cite{Effective_mass_STO} where $m$ is free electron mass. Within the hydrogenic theory of
shallow donors, the donor Bohr radius is equal to  $\kappa b$, where $ b = \hbar^2/m^* e^2 \simeq 0.35 ~\mathrm{\AA}$, $e$ is the electron charge, and $\kappa$ is dielectric constant of the material. At room temperature when $\kappa = 350$, the Bohr radius  $\kappa b = \mathrm{123 ~ \AA}$ is so large that the Mott criterion for the metal-insulator transition in doped semiconductors $N_c b^3 =0.02/\kappa^3$ leads to a very small critical concentration $N_c = 1\cdot 10^{16} ~\mathrm{cm^{-3}}$. At helium temperatures $\kappa=2 \cdot 10^{4}$ and $N_c =6 \cdot 10^{10} ~ \mathrm{cm^{-3}} $. Thus, at the experimentally relevant
concentration of donors  $N>10^{17} ~\mathrm{cm^{-3}}$, we are dealing with a heavily doped semiconductor in which the Fermi energy lies in the conduction band of $\mathrm{SrTiO_3}$. On the other hand,  due to the relatively high effective mass the bulk Fermi energy $\varepsilon_F$ is  smaller than  the bending energy of the conduction band bottom  near the interface (see Figs. \ref{fig:accumultion} and~\ref{fig:depletion}). For example, for $N =10^{18} \mathrm{cm^{-3}}$, the Fermi energy calculated from the bottom of the conduction band is  $\varepsilon_F \simeq 4~\mathrm{meV},$ which can be up to $100$ times smaller than the bending energy of the conduction band bottom in an accumulation layer for $\mathrm{GdTiO_3/SrTiO_3}$. Therefore,  we assume below that the Fermi energy  coincides with the bottom of the conduction band.

We are interested in accumulation, inversion and depletion layers near an interface of $\mathrm{SrTiO_3}$. We consider the case when the axis $x$ is directed perpendicular to the interface  (plane $x=0$) and lies along the [100] axis of a cubic crystal of $\mathrm{SrTiO_3}$. (In fact, $\mathrm{SrTiO_3}$  changes symmetry  from cubic to tetragonal at  $T \simeq 110 K$, but the distortion is small \cite{STO_lattice} and can be neglected). An external electric field $D_0$ applied from the left (see Figs. \ref{fig:accumultion},~\ref{fig:inversion}, \ref{fig:depletion})  is directed along the $x$ axis. In that case the problem is effectively one-dimensional. If the charge density  is denoted by $\rho(x)$, then  the potential depth profile $\varphi(x)$ in the system is determined by the  equations: 

\begin{eqnarray}
  \label{eq:Gauss}
  \frac{d D}{dx}   = 4 \pi  \rho ,  ~  D =E+4\pi P , ~   \frac {d \varphi}{d x}  = -E ,
\end{eqnarray}
where $D(x), E(x), P(x)$ are electric induction, electric field and  polarization in $\mathrm{SrTiO_3}$.  Equations (\ref{eq:Gauss}) should be solved with proper boundary conditions. For example, for an accumulation layer the boundary conditions are $D(0)=D_0$ and $\varphi(\infty)=0$.

To solve the system (\ref{eq:Gauss})  one needs to know two material relationships  $E(P)$ and $\rho(\varphi)$. Let us start from the lattice dielectric response $E(P)$. $\mathrm{SrTiO_3}$ is well known as a quantum paraelectric, where the onset of ferroelectric order is suppressed by quantum fluctuations \cite{Ferroelectricity_by18O}. 

A powerful approach to describe the properties  of ferroelectric-like materials  is based on the Landau-Ginzburg   theory.  For a continuous second-order phase transition the Landau-Ginzburg expression of the free energy density $F$   is represented as a power series expansion with respect to the polarization $P$:
\begin{equation}
  \label{eq:F}
  F= F_0 +\frac{\tau}{2} P^2 + \frac{1}{4} A \frac{1}{P_0^2} P^4  -E P, 
\end{equation}
where $F_0$ stands for the free energy density at $P = 0$ and $\tau$ is the inverse susceptibility $\tau=4\pi/(\kappa-1) \simeq 4\pi/\kappa$. In this work $0<\tau \ll 1$, $P_0=e/a^2$ is the characteristic polarization and $a \simeq \mathrm{3.9 ~\AA}$ \cite{STO_lattice} is the lattice constant. The coefficient A  describes the non-linear dielectric response. Analyzing the available data \cite{Berg_Schottky_STO,  JJAP.36.L390, Shotky_Landau_starnge_formula,STO_Shottky_Mikheev} in Section \ref{sec:depletion}  we find  values of $A$ between 0.5 and 1.5. For all estimates below we use $A=0.8$ following from Ref. \cite{STO_Shottky_Mikheev}. The last term of Eq. (\ref{eq:F}) is responsible for the interaction between the polarization and the electric field $E$. In general $F$  depends on the components of the vector $P$, but in the chosen geometry  the problem is one-dimensional, and all vectors are directed along the $x$ axis.
The crystal polarization $P$ is determined by minimizing the free energy density $F$ in the presence of the electric field $E$, $\delta F/ \delta P=0$. This condition  relates $E$ and $P$,

\begin{equation}
  \label{eq:electric_field_definition}
 E = \frac{4\pi}{\kappa} P + \frac{A}{P_0^2} P^3.
\end{equation}
We note that $E \ll 4\pi P$ and thus $D=E+4\pi P \simeq 4\pi P$. The electric field $D_c$ at which the transition from linear to nonlinear dielectric response  occurs can be found by equating the first and second terms in the expression (\ref{eq:electric_field_definition}):
\begin{equation}
  \label{eq:Dc}
D_c = P_0 \sqrt{\frac{(4\pi)^3}{\kappa A}}.
\end{equation}

If $D \ll D_c $  the dielectric response of $\mathrm{SrTiO_3}$ is linear and one can use the simplified expression for the electric field:  
\begin{equation}
  \label{eq:electric_field_small}
 E =  \frac{D}{\kappa}.
\end{equation}

For $D \gg D_c $  the dielectric response of $\mathrm{SrTiO_3}$ is nonlinear and one must instead use the expression:

\begin{equation}
  \label{eq:electric_field_high}
E= \frac{A}{(4\pi)^3 P_0^2} D^3.
\end{equation}
Next one should specify $\rho(\varphi)$, which depends on the specific device of interest. 

\section{Accumulation  layer: Theory}
\label{sec:accumulation}
\begin{figure}
\includegraphics[width=0.8\linewidth]{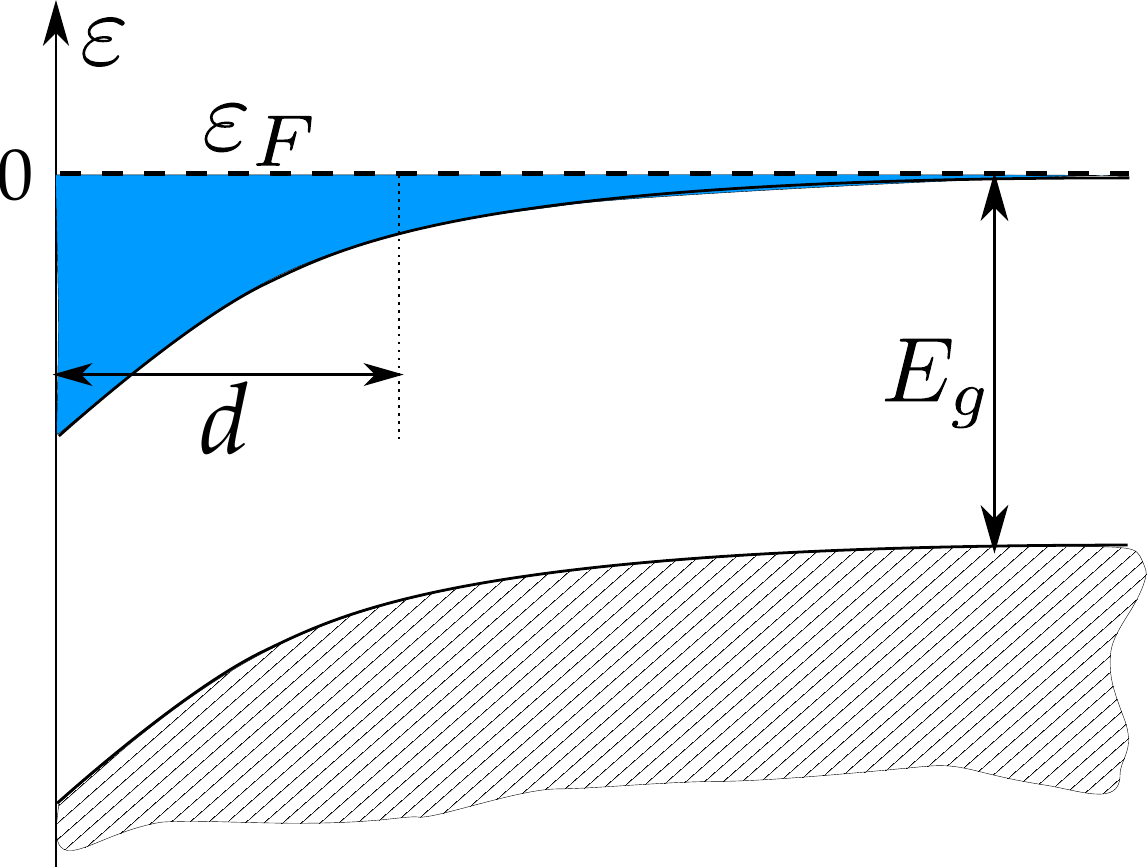}
\caption{(Color  online) Schematic energy diagram of an accumulation layer in an $n$-doped semiconductor with  band gap $E_g$. Electrons (blue region) are attracted by an external electric field $D_0$. The characteristic width of the electron gas is $d$. In the bulk of  $\mathrm{SrTiO_3}$  the Fermi level   $\varepsilon_F $ is near the bottom of the conduction band  (plotted by the dashed line)}
\label{fig:accumultion}
\end{figure}

In  an accumulation  layer  the external electric field $D_0$  attracts electrons with a three-dimensional concentration $n(x)$ (see Fig. \ref{fig:accumultion}). Our goal is to find  the electron depth profile $n(x)$ and its characteristic width $d$. 

Due to electric neutrality the number of accumulated electrons has to compensate the external field $D_0$, i.e., 
\begin{equation}
  \label{eq:neutrality}
  4 \pi e \int \limits_0^{\infty} n(x) dx = D_0.
\end{equation}
To take into account the electron screening of the external field  we  use the Thomas-Fermi approach \cite{Thomas_Fermi}  in which the electron concentration $n(x)$ and self-consistent potential profile $\varphi(x)$ are related as $e\varphi(x)+\mu(x)=\varepsilon_F=0$, where 
\begin{equation}
  \label{eq:TF}
  \mu(x)= (3\pi^2)^{2/3}\frac{\hbar^2}{2 m} [n(x)]^{2/3}
\end{equation}
is the chemical potential of the electron gas. Thus, one can obtain the solution of Eqs. \eqref{eq:Gauss} by replacing  $\rho(x)$ with $e n(x)$ and using relations \eqref{eq:electric_field_small} and \eqref{eq:electric_field_high}. For a linear dielectric response we obtain the equation for the potential: 
\begin{equation}
  \label{eq:potential_equation_linear}
\frac{d^2 }{dx^2} \left(\frac{\varphi}{e/b}\right) = \frac{2^{3/2}}{3\pi^2} \frac{1}{b^2} \frac{1}{\kappa} \left(\frac{\varphi}{e/b}\right)^{3/2}.  
\end{equation}
We use the boundary condition $\varphi=0$  at $x \rightarrow \infty$ and get the solution:
\begin{equation}
  \label{eq:potential_linear}
  \varphi(x)=  C_1 \frac{e}{b} \kappa^2   \left(\frac{b}{x+d} \right)^4,
\end{equation}
\begin{equation}
  \label{eq:conventration_linear}
  n(x)= C_2 \frac{1}{b^3}   \kappa^3 \left(\frac{b}{x+d}\right)^{6},
\end{equation} 
where  $C_1=(225/8)\pi^2 \simeq 278$ and $C_2=1125 \pi /8 \simeq 442 $. ($\varphi(0)$ was derived equivalently in Ref. \onlinecite{Stratton_field_emission} as a work function reduction for $\mathrm{GaAs}$.) For a nonlinear dielectric response we obtain the equation for the potential: 
\begin{equation}
  \label{eq:potential_nonlinear_equation}
\frac{d}{dx} \left[\left(\frac{d}{dx} \frac{\varphi}{e/b}\right)^{1/3}\right] = \frac{2^{3/2}}{3\pi^2} \frac{1}{b^{4/3}} A^{1/3}\left(\frac{e/b^2}{P_0}\right)^{2/3} \left(\frac{\varphi}{e/b}\right)^{3/2}.  
\end{equation}
With the same boundary condition we get the solution:
\begin{equation}
  \label{eq:potential_nonlinear}
  \varphi(x)=  C_3 \frac{e}{b} \left(\frac{b}{a}\right)^{8/7} \frac{1}{A^{2/7}} \left(\frac{b}{x+d} \right)^{8/7},
\end{equation}
\begin{equation}
  \label{eq:conventration_nonlinear}
  n(x)=C_4 \frac{1}{b^3} \left(\frac{b}{a}\right)^{12/7} \frac{1}{A^{3/7}}  \left(\frac{b}{x+d}\right)^{12/7},
\end{equation} 
where $C_3= [5^6 3^6 \pi^{12}/(7^8 2^3)]^{1/7} \simeq 5.8,$ $C_4= [5^{9} 3^{2} \pi^{4} 2^6/7^{12}]^{1/7} \simeq 1.3.$

The characteristic length $d$ can be obtained using the neutrality condition (see Eq. (\ref{eq:neutrality})). For a linear dielectric response this gives:

\begin{equation}
  \label{eq:d_small}
  d= C_5 b \left(\frac{a}{b}\right)^{2/5} \kappa^{3/5} \left(\frac{e/a^2}{D_0}\right)^{1/5},
\end{equation}
where $C_5=[\pi^2 225/2 ]^{1/5} \simeq 4$. For a nonlinear dielectric response:

\begin{equation}
  \label{eq:d_STO}
 d= C_6 b \left(\frac{a}{b}\right)^{2/5} \left (\frac{e/a^2}{D_0}\right)^{7/5} \frac{1}{A^{3/5}},
\end{equation}
where $C_6 = (16/7) (5^2 3^2 \pi^{11})^{1/5} \simeq 84$. The electric field $D_c$ at which the transition from linear to nonlinear dielectric response  occurs can be found from equating Eqs.  \eqref{eq:d_small} and   \eqref{eq:d_STO}. This gives
\begin{equation}
  \label{eq:4}
D_c = \frac{C_7}{\sqrt{A}} \frac{e}{a^2}\sqrt{\frac{1}{\kappa}},   
\end{equation}
where $C_7=(2^{21} \pi^9/7^5)^{1/6} \simeq 12$, consistent with Eq. (\ref{eq:Dc}). For $\mathrm{SrTiO_3}$, the critical field $D_c$ depends on temperature:  $D_c \simeq 0.1 e/a^2$ for helium temperature and $D_c \simeq 0.7 e/a^2$ for room temperature. 

The three dimensional concentration profile  $n(x)$ for the nonlinear dielectric response  Eq. (\ref{eq:conventration_nonlinear}) is the main result of our paper. Note that $n(x)$ has a very long tail with a weak  $12/7$ power law dependence, which may lead to some arbitrariness in measurements of the width of the electron gas. Indeed, only 39\% of electrons are located within the distance $0< x < d$ near the interface and 68\% of electrons are located within $0<x < 4 d$. In the calculation above we used a space-continuous model. Actually,  along the [100] axis $\mathrm{SrTiO_3}$  is composed of alternating $\mathrm{TiO_2}$ and $\mathrm{SrO}$ layers. The conduction band of $\mathrm{SrTiO_3}$ corresponds to the bands composed of mainly $3d$ orbitals of $\mathrm{Ti}$. Integrating $n(x)$ over each lattice cell in Table  (\ref{tbl:distribution}) we get a percentage of electrons in each of the 10 first $\mathrm{TiO_2}$ layers of $\mathrm{SrTiO_3}$ for the case $D=2\pi e/a^2$.

\begin{table}
\caption{\label{tbl:distribution} Percentage of electrons   in the $\mathrm{TiO_2}$ layer $\textnumero M$ of $\mathrm{SrTiO_3}$ for  $D_0=2\pi e/a^2$, corresponding to a total surface density of $0.5 e /a^2$ }
\begin{ruledtabular}
\begin{tabular}{ c   c c c c c c c c c c}
M & 1 & 2& 3&4&5&6&7&8&9&10 \\ \hline
Percent &27.9&14.4&9.0&6.2&4.6&3.5&2.8&2.3&1.9&1.6 \\ 
\end{tabular}
\end{ruledtabular}
\end{table}

One can see from Eqs. (\ref{eq:conventration_linear}) and (\ref{eq:conventration_nonlinear})  that the tails of the electron depth profiles $n(x)$ at $x \gg d$ do not depend on $D_0$ and behave like $$C_2 \frac{1}{b^{3}} \kappa^3 \left(\frac{b}{x}\right)^6$$ and $$ C_4 \frac{1}{b^3} \left(\frac{b}{a} \right)^{12/7} \frac{1}{A^{3/7}} \left(\frac{b}{x}\right)^{12/7}$$ for linear and nonlinear dielectric responses, respectively. Even for $D_0 \gg D_c$,  when the electron distribution $n(x)$ at moderately large  $x$ is described by dependence (\ref{eq:conventration_nonlinear}), at very large distances the polarization becomes smaller and the linear dielectric response takes over so that the $n(x)$ dependence switches from Eq. (\ref{eq:conventration_nonlinear}) to Eq. (\ref{eq:conventration_linear}). This happens at the distance $$x_0=b \left(\frac{C_2}{C_4}\right)^{7/30}  A^{1/10} \left(\frac{a}{b}\right)^{2/5} \kappa^{7/10}$$ ( $x_0 =360~\mathrm{nm}$ and $20~\mathrm{nm}$ for helium and room temperature respectively). Thus, the tail of $n(x)$ is universal. For small $D_0<D_c$ the tail has the form $n(x) \propto x^{-6}$. For $D_0>D_c$ it has the form  $n(x) \propto x^{-12/7}$ for $x<x_0$ and $n(x) \propto x^{-6}$ for $x>x_0$. 

On the other hand, one has to remember  that  our theory is correct only when $n(x)$  is larger than the concentration of donors in the bulk of the material.

Let us verify whether the Thomas-Fermi approximation is applicable, i.e.,  $k_F d \gg 1$. Here $k_F = (3\pi^2)^{1/3}n(0)^{1/3}$ is the wavevector of an electron at the Fermi level. For $D_0 \ll D_c$

\begin{equation}
  \label{eq:linear_kfd}
  k_Fd = C_8 \kappa^{2/5} \left(\frac{b}{a}\right)^{2/5} \left(\frac{D_0}{e/a^2}\right)^{1/5},
\end{equation}
while  for $D_0 \gg D_c$
\begin{equation}
  \label{eq:cubic_kfd}
  k_Fd = C_9 \frac{1}{A^{2/5}} \left(\frac{b}{a}\right)^{2/5}  \left(\frac{e/a^2}{D_0}\right)^{3/5},
\end{equation}
 where $C_8= (5^33^3\pi^3/2^4)^{1/5} \simeq 6$, $C_9=4/7 (15 \pi^3)^{3/5} \simeq 23.$ One can see that  $k_F d>1$ in the range of $2 \cdot 10^{-7} e/a^2 < D_0< 40 ~e/a^2 $ for room temperature. For lower temperatures this interval is even  larger. Thus, the Thomas-Fermi approximation is applicable for practically  all  reasonable electric fields $D_0$.  \footnote{So far we considered only two limiting cases: the linear and nonlinear dielectric responses, which are correct for $D_0 \ll D_c$ and $D_0 \gg D_c$ respectively. In fact one can analytically solve Eqs. (\ref{eq:Gauss}) with  Eq. (\ref{eq:electric_field_definition}) assuming that $D = 4\pi P$. For example, in Ref. \onlinecite{Gureev_ferroelectric_electrons}  the polarization profile of a charged ``head-to-head''  domain wall in a ferroelectric  was calculated. In the ferroelectric phase, the spontaneous polarization in the  domain is approximately equal to our $D_c$. Therefore a charged domain wall,  considered in Ref. \onlinecite{Gureev_ferroelectric_electrons}, should be compared with our accumulation layer at the crossover value of the electric field, $D_0=D_c$. Indeed, our results Eqs. (\ref{eq:conventration_linear}), (\ref{eq:d_small}) taken at $D_0=D_c$  agree with those for the charged domain wall.}


\section{Accumulation layer: Comparison with experimental data}
\label{sec:experiment}
It is widely believed that an electron gas  emerges near polar/non-polar interfaces such as $\mathrm{LaAlO_3/SrTiO_3},$ $\mathrm{GdTiO_3/SrTiO_3}$ \cite{Stemmer_GdTO,Stemmer_concentration_interface}, $\mathrm{LaVO_3/SrTiO_3}$ \cite{LaVO_STO}, $\mathrm{NdAlO_3/SrTiO_3}$, $\mathrm{PrAlO_3/SrTiO_3}$, $\mathrm{NdGaO_3/SrTiO_3}$ \cite{different_polar_STO}, $\mathrm{LaGaO_3/SrTiO_3}$ \cite{LaGaO_STO}  and $\mathrm{LaTiO_3/SrTiO_3}$ \cite{LaTO_STO} due to the polar catastrophe \cite{polar-catastrophe_2006}. For a large enough thickness of the polar crystal  \cite{abinitio_STO_2} the interface   electron surface charge density 
\begin{equation}
  \label{eq:sigma_def}
  \sigma = e \int \limits_0^{\infty} n(x)dx
\end{equation}
is equal to $0.5 e/a^2$, which corresponds to   $D_0 =2\pi e/a^2$ [see Eq. (\ref{eq:neutrality})]. For any temperature this field is much larger then the critical field $D_c$ and Eq. (\ref{eq:cubic_kfd}) gives $k_Fd = 3$. Thus, we arrive at Eq. (\ref{eq:conventration_nonlinear}) for the electron concentration $n(x)$ and  Eq. (\ref{eq:d_STO}) shows that $d \simeq 6.7~ \mathrm{\AA}$.  In this case 68\%  of electrons are located within $2.7~ \mathrm{nm}$.  This result agrees  with   experimental estimates of the width of an electron gas near the $\mathrm{GdTiO_3/SrTiO_3}$ interface \cite{Stemmer_GdTO}. It also agrees with experimental data \cite{spinel_STO} for  the $\mathrm{\gamma \textendash Al_2O_3/SrTiO_3}$ interface where a 2DEG is formed due to the formation of oxygen vacancies on $\mathrm{SrTiO_3}$. In the case of $\delta$-doped $\mathrm{SrTiO_3}$ with one layer of La \cite{Ohtomo_Hwang_LaTiO_STO} one has two accumulation layers each with $\sigma=0.5 e/a^2$ and a similar width $d$.

To see how important the nonlinear dielectric response is, one can compare its prediction to the one obtained by assuming the response to be linear, given by Eq. (\ref{eq:d_small}). For $D=2\pi e/a^2$ and helium temperature the linear dielectric response gives $d=167 ~\mathrm{nm}$. A similar result was obtained in Ref. \onlinecite{wrong_estimation}, where the nonlinear dielectric response was not taken into account.

For the  $\mathrm{LaAlO_3 /SrTiO_3}$ interface the number of electrons accumulated is apparently smaller than what the ``polar catastrophe'' scenario predicts. For example, only $\simeq 10\% $ of the electrons are seen in Hall measurements \cite{10_percent,10_percent_2, 10_percent_VRH}. In order to describe the electron concentration $n(x)$ for such a surface charge density, $\sigma = 0.05 e/a^2$, one can still use  Eq. (\ref{eq:conventration_nonlinear}) and Eq. (\ref{eq:d_STO}). As a result, we arrive at a much larger value of  $d \simeq 17 ~\mathrm{nm}$.

\begin{figure}
\includegraphics[width=1\linewidth]{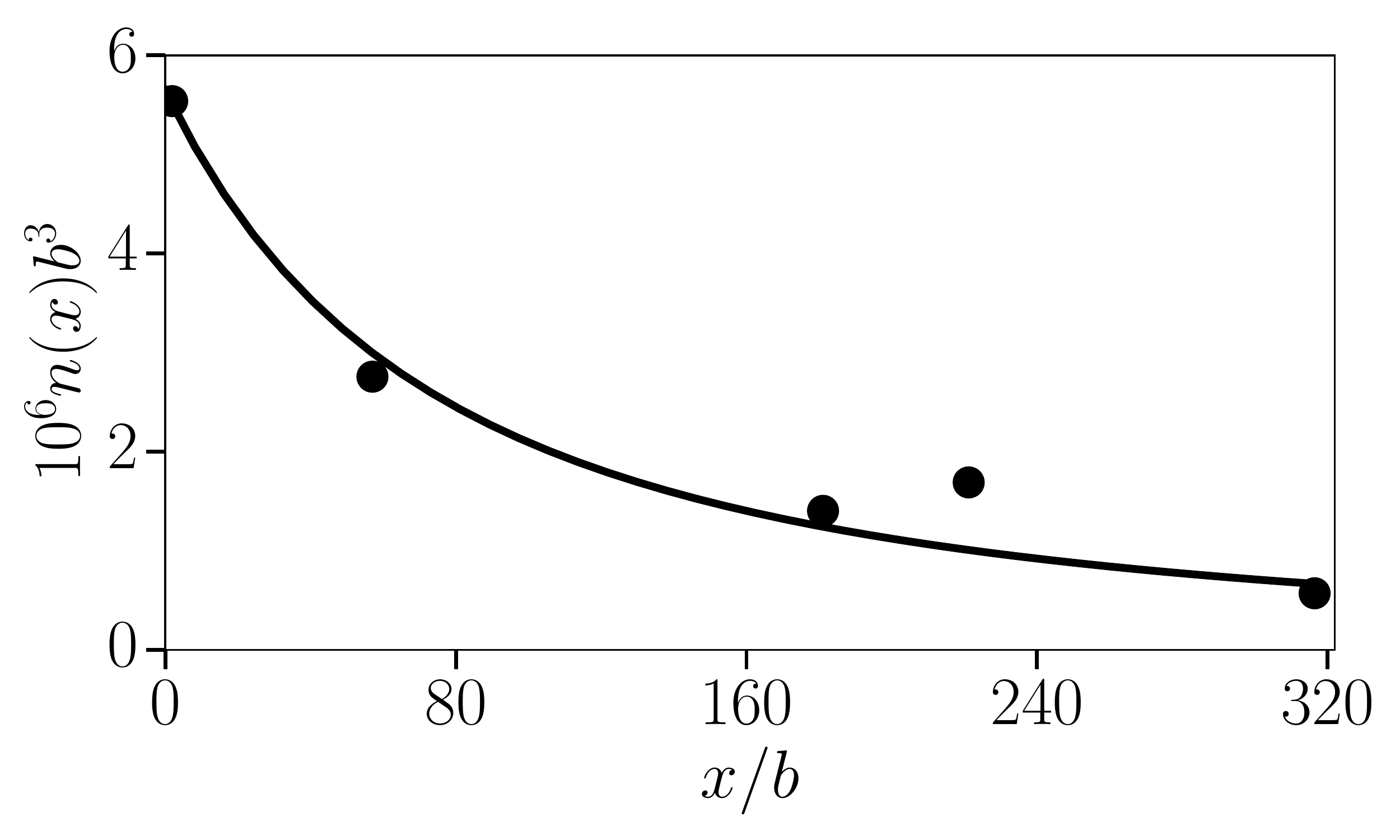}
\caption{ The experimental distribution of electrons $n(x)$ from Ref. \onlinecite{LAO_STO_Berreman}  is shown by circles  in convenient dimensionless units. Fitting by Eq.~(\ref{eq:conventration_nonlinear}) is shown by the solid line. The fitting parameter is $\sigma=0.13 e/a^2$, which gives $d \simeq 142 b \simeq 5~ \mathrm{nm}$.}
\label{fig:experiment}
\end{figure}

We test our theory for the functional shape of $n(x)$ by comparing to  experimental data for the electron distribution $n(x)$ near the interface of $\mathrm{LaAlO_3 /SrTiO_3}$ at temperature $\simeq  10~ \mathrm{K}$   \cite{LAO_STO_Berreman} (see Fig. \ref{fig:experiment}). For such small temperatures the critical field $D_c$ is small and we fit the experimental data by  Eq. (\ref{eq:conventration_nonlinear}) with $\sigma=D_0/4\pi$ as a fitting parameter and get  $\sigma= 0.12 e/a^2$ with $d \simeq 142 b \simeq 5~ \mathrm{nm}$.  Figure \ref{fig:experiment} shows satisfactory agreement between the data and the shape of $n(x)$ described by Eq.~(\ref{eq:conventration_nonlinear}). 

\begin{figure}
\includegraphics[width=1\linewidth]{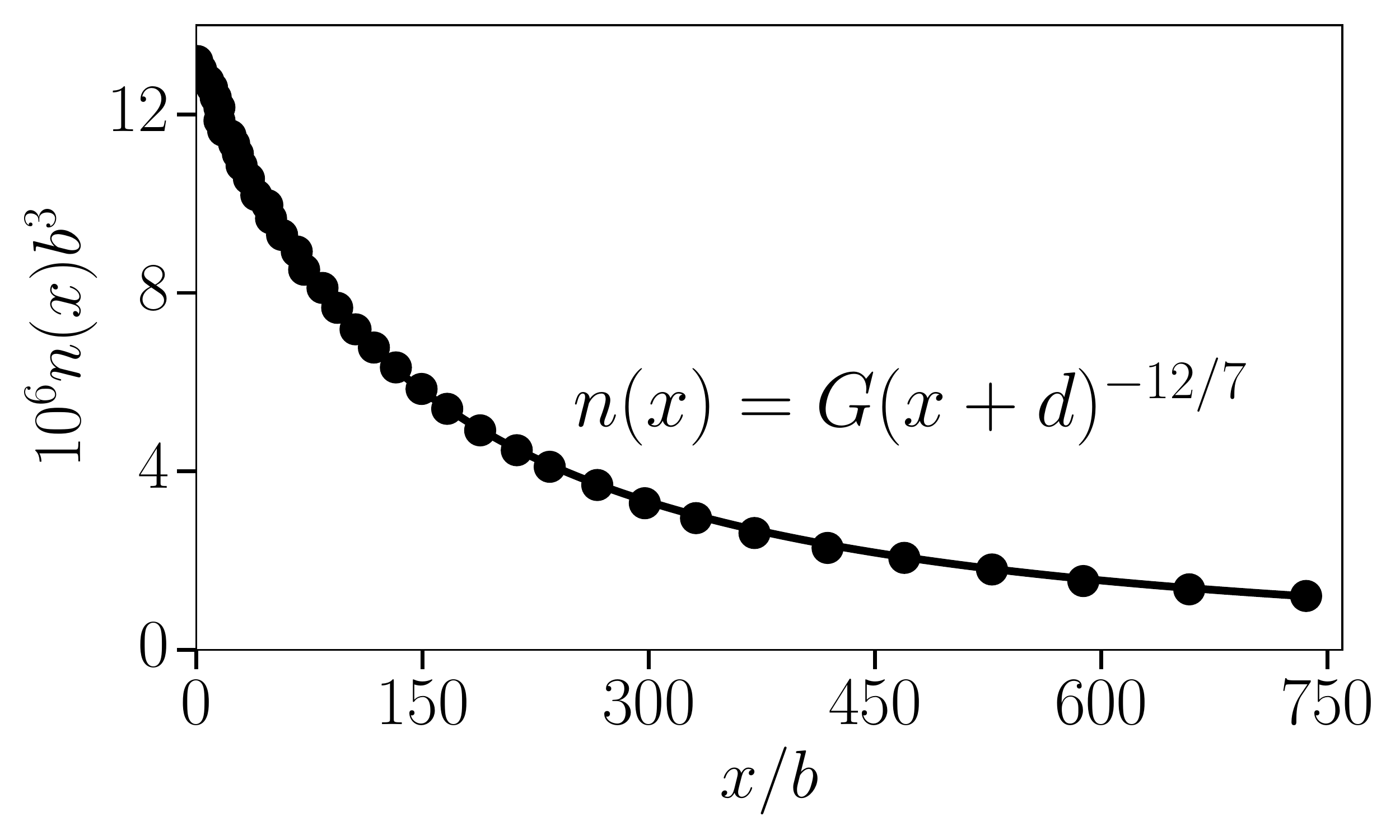}
\caption{ The experimental distribution of electrons $n(x)$ (circles) from Ref. \onlinecite{Hwang_PL}  obtained by time-resolved photoluminescence in convenient dimensionless units. Fitting with Eq.~(\ref{eq:new_n_x}) is shown by the solid line. The fitting parameters are $G=0.16$ and  $d = 250 b \simeq 9~ \mathrm{nm}.$}
\label{fig:Hwang_PL}
\end{figure}

Let us now dwell upon the experimental  data for $n(x)$ from Ref. \onlinecite{Hwang_PL}. The data are obtained  via    time-resolved photoluminescence spectroscopy of the  $ \mathrm{LaAlO_3/SrTiO_3}$ interface, where interface-induced electrons radiatively recombine with photoexcited holes in $\mathrm{SrTiO_3}$. Following the assumption from Ref. \onlinecite{Hwang_PL} that the photoexcited holes are immobile, the concentration of holes $p$ decays with time $t$ according to the equation:
\begin{equation}
  \label{eq:holes_dynamic}
  \frac{dp}{dt} = -r_1 p - r_2 \left[n(x)\right]^2 p,
\end{equation}
where $r_1$ is the hole trapping rate and $r_2$ is the three carrier non-radiative Auger recombination coefficient. The authors of Ref.~\onlinecite{Hwang_PL} used the decay of  the photoluminescence  intensity  to obtain $n(x)$ in $\mathrm{LaAlO_3/SrTiO_3}$ with the help of the coefficients $r_1,r_2$ from Ref. \onlinecite{Auger_coefficients_PL_STO}. The resulting $n(x)$  is shown in Fig. \ref{fig:Hwang_PL} by filled circles.  We fit this data by the equation
\begin{equation}
  \label{eq:new_n_x}
  n(x)=\frac{G}{(x+d)^{12/7}},
\end{equation}
which is similar to Eq. (\ref{eq:conventration_nonlinear}), but  $G$ is a fitting parameter independent of $d$. We see from Fig.~\ref{fig:Hwang_PL} that this fit is good, however, the parameter $G$  is seven times larger than the parameter $C_4 (b/a)^{12/7}/A^{3/7}$ entering Eq. (\ref{eq:conventration_nonlinear}). This can be explained by a 50-fold increase of  the Auger coefficient $r_2$ of Eq. (\ref{eq:holes_dynamic}) near the interface. A similar surface effect was observed in  Ref. \onlinecite{Auger_coefficients_PL_nano_STO} for $\mathrm{SrTiO_3}$ nanocrystals, where  $r_2$ is almost 150 times larger than for bulk $\mathrm{SrTiO_3}$.  

From the fit we get $d=250b \simeq 9~\mathrm{ nm}$, which corresponds to an electron surface charge density  $\sigma=D_0/4\pi =0.08 e/a^2$ in agreement with other data \cite{10_percent,10_percent_2,LAO_STO_Berreman}. One can check that this result is self-consistent, i.e. for such $\sigma$ one can use Eq. (\ref{eq:conventration_nonlinear}) for the fitting of experimental data, because $k_Fd > 1$ (see Eq. (\ref{eq:cubic_kfd})) and $D_0>D_c$ (see Eq. (\ref{eq:4})).

\section{Quantum capacitance of an accumulation layer}
\label{sec:quantum}

In this section we address the capacitance of an accumulation layer, for example in the double junction $\mathrm{Metal/GdTiO_3/SrTiO_3}$. If the width of insulating $\mathrm{GdTiO_3}$ layer $L \gg d$ we may view an accumulation layer as a conducting two-dimensional gas (2DEG). One can apply a positive voltage $V$ between the metal and the 2DEG and measure the additional charge per unit area $\sigma'$ and $-\sigma'$, which are induced on the metal and in the 2DEG, respectively. The capacitance per unit area of junction is $C=d\sigma'/dV$. If we imagine that the 2DEG is a perfect metal, the additional charge $-\sigma'$ resides exactly in the plane of the $\mathrm{GdTiO_3/SrTiO_3}$ junction and  the capacitance is equal to the geometric capacitance $C=C_G = \kappa_G /4\pi L$, where $\kappa_G=30$ is the dielectric constant of the $\mathrm{GdTiO_3}$ layer. Actually, the accumulation layer is not a perfect metal so that an additional negative charge $-\sigma'$ is distributed in a layer of finite width $d$. As a result $C^{-1} = C_G^{-1} + C_q^{-1} $, where  $C_q$ is called the quantum capacitance~\cite{Luryi}. Quantum capacitance is broadly studied for many 2DEGs such as silicon MOSFETs ~\cite{quantum_capacitance_Kravchenko}, $\mathrm{GaAs/Ga_{x}Al_{1-x}As}$ hetero-structures~ \cite{PhysRevLett.68.674} and graphene~\cite{quantum_capacitance_graphene}. 

Concentrating on the case of $D_0 \gg D_c$ and using Eqs. (\ref{eq:potential_nonlinear_equation}) and (\ref{eq:d_STO}) for  the potential difference between $x=0$ and $x=\infty$ we have at $\sigma'=0$:

\begin{equation}
  \label{eq:phi_0}
  \varphi(0)=\frac{C_3}{C_6^{8/7}} \frac{e}{b}  A^{2/5} \left(\frac{b}{a}\right)^{8/5} \left(\frac{D_0}{e/a^2}\right)^{8/5}.
\end{equation}
After the transfer of charge $-\sigma'$ from the metal  to the electron gas  the potential changes as:
\begin{equation}
  \label{eq:phi_after}
  \Delta \varphi=\frac{C_3}{C_6^{8/7}}  \frac{e}{b}  A^{2/5} \left(\frac{b}{a}\right)^{8/5} \left[\left(\frac{D_0+4\pi \sigma'}{e/a^2}\right)^{8/5} - \left(\frac{D_0}{e/a^2}\right)^{8/5} \right].
\end{equation}

Assuming that  $D_0 \gg 4\pi \sigma'$, linearizing $\Delta \varphi$ with respect of $\sigma'$ and adding the voltage drop  across the $\mathrm{GdTiO_3}$ layer we get
$$
V=\frac{e}{b} \frac{8}{5} \frac{C_3}{C_6^{8/7}} A^{2/5} \left(\frac{b}{a}\right)^{8/5} \frac{4\pi \sigma'}{e/a^2} \left(\frac{D_0}{e/a^2}\right)^{3/5} + \frac{4\pi L \sigma'}{\kappa_G}.
$$

Using Eq. (\ref{eq:d_STO}) for $d$ we can write for the total capacitance:
\begin{equation}
  \label{eq:capacitance_quantum}
  C^{-1} = \frac{7}{5}\frac{4\pi d}{\kappa_{eff}} + \frac{4\pi L}{\kappa_G},
\end{equation}
where $$\kappa_{eff}= \frac{(4\pi)^3}{A (D_0/(e/a^2))^2}$$ is the effective nonlinear dielectric constant $D/E$ at $x=0$ (see Eq. (\ref{eq:electric_field_high})) . The first term of Eq. (\ref{eq:capacitance_quantum}) is the inverse quantum capacitance $C_q^{-1}$, and the last term is the inverse geometric capacitance $C_G^{-1}$.

Due to the polar catastrophe in $\mathrm{GdTiO_3}$ we get $D_0=2\pi e/a^2$ and $\kappa_{eff} = 63$. The ratio of inverse capacitances is $$\frac{C_{q}^{-1}}{C_{G}^{-1}}=\frac{7}{5} \frac{ \kappa_G}{\kappa_{eff}} \frac{d}{L}.$$ For $L=40~\mathrm{\AA}$ we get $C_{q}^{-1}/C_{G}^{-1}=0.1$. One can extend our calculation to the  double junction $\mathrm{SrTiO_3/GdTiO_3/SrTiO_3}$. In that case the inverse quantum capacitance is the sum of inverse quantum capacitances of both junctions.

\section{From accumulation to inversion layer}
\label{sec:inversion}

\begin{figure}
\includegraphics[width=0.8\linewidth]{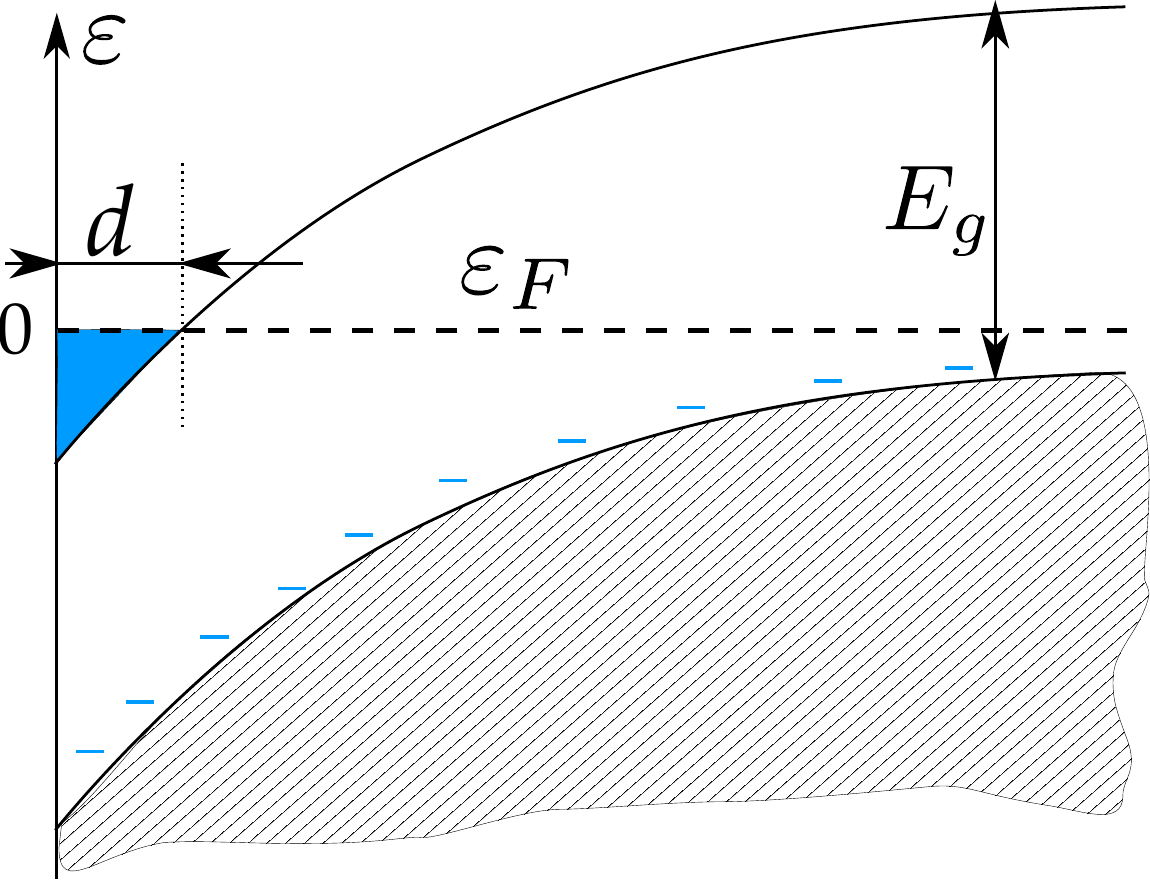}
\caption{(Color  online) Inversion layer of a $p$-type semiconductor. Holes  have been forced away by an external electric field $D_0$. The field is so strong that it attracts mobile electrons (blue region) in the inversion layer with width $d$ and total surface charge density $\sigma$.  Negatively charged acceptors are shown by blue minuses. The Fermi level $\varepsilon_F$ is plotted by a dashed line.}
\label{fig:inversion}
\end{figure}

In the previous section we considered an accumulation layer and showed that it can be described
by the Thomas-Fermi approximation, when the Fermi level is near the bottom of the conduction band. But the Fermi level can be moved into the gap, for example, by a back gate.  In a $p$-type inversion layer in $\mathrm{SrTiO_3}$ (see Fig. \ref{fig:inversion}) the Fermi level can be even deeper, at the top of the bulk valence band. In both cases, the charge of electrons compensates only a fraction of the external electric field $D_0$, i. e., Eq. (\ref{eq:neutrality}) is violated. For example, the rest of the negative charge in the inversion layer is provided by negative acceptors  [see Fig. \ref{fig:inversion}] and the   electron surface  charge density  $\sigma < D_0/(4 \pi)$.
To calculate  the width of the electron density profile $d(\sigma, D_0)$ below we  use a scaling approach, i.e. we neglect all numerical coefficients.

\begin{figure}
\includegraphics[width=\linewidth]{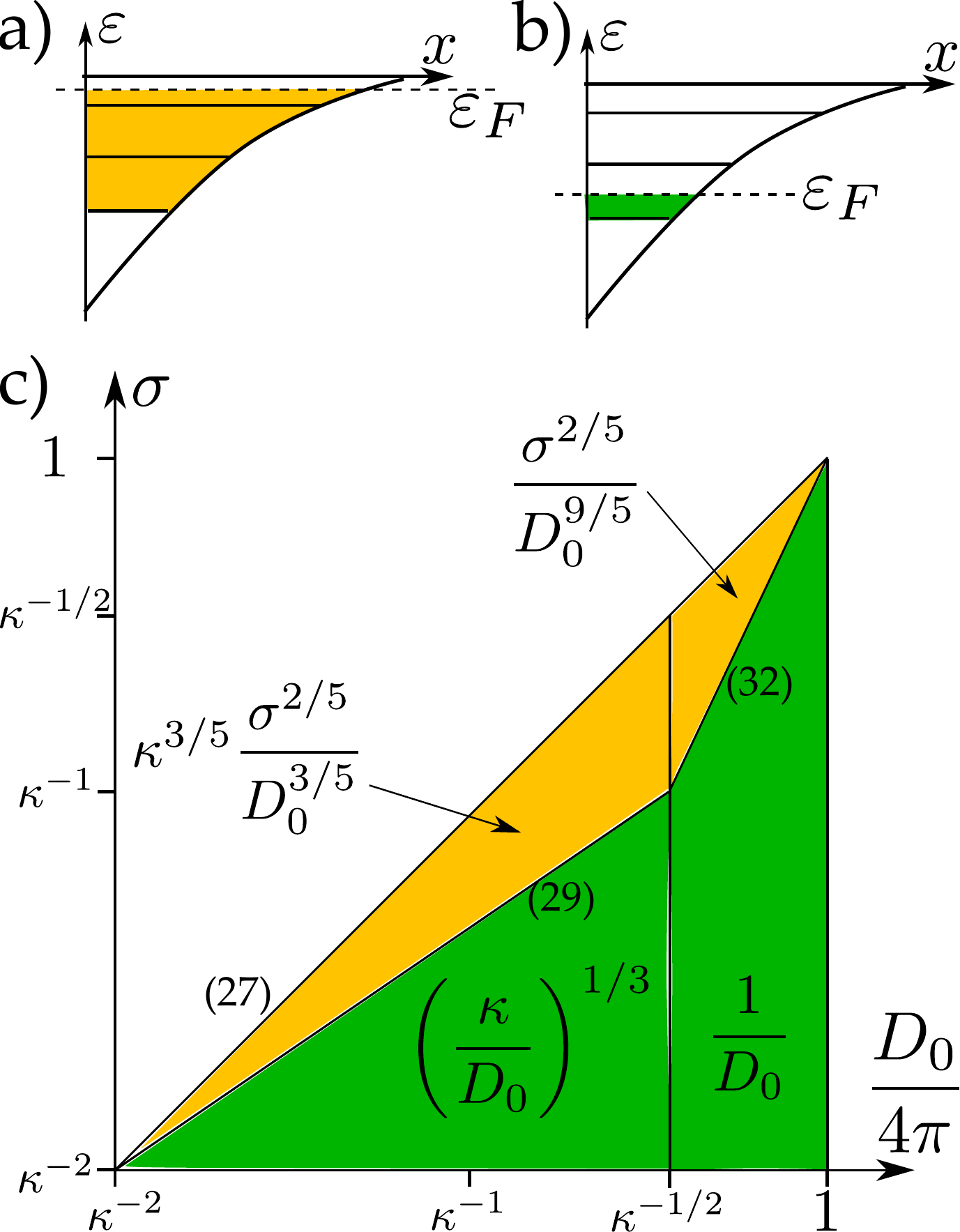}
\caption{(Color online) Schematic  energy diagram for conduction band electrons   in the triangular potential well created by an external electric field $D_0$  for two different cases: a) 3DEG, when many sub-bands are filled by electrons and b) 2DEG, when only the lowest sub-band is partially filled. This filling  is shown by light yellow and dark green, respectively. c) Schematic phase diagram for the dependence of the width of electron gas  $d$ in units of $b$ on electron surface charge density  $\sigma$ and external electric field $D_0$ both in the units $e/b^2$  for $\mathrm{SrTiO_3}$ with dielectric constant $\kappa$. The four domains  of this diagram  correspond to four different  formulas for $d$.  If the electric field is small,  $D_0<1/\sqrt{\kappa}$, the  linear relation between electric field and electric induction, Eq. (\ref{eq:electric_field_small}), is applicable. For $D_0>1/ \sqrt{\kappa}$ Eq.  (\ref{eq:electric_field_small}) is replaced by the nonlinear relation Eq. (\ref{eq:electric_field_high}). The light and dark grey  (yellow and green in color) regions  distinguish 3DEG and 2DEG cases. The border lines between domains  are  marked by numbers of corresponding equations (\ref{eq:critical_trivial}),~(\ref{eq:critical1}), ~(\ref{eq:critical2}).}
\label{fig:phase_1}
\end{figure}

If the surface electron  density  is high enough $\sigma \gg e/d^2 $, the  electron gas is three-dimensional (3DEG), and one can use the Thomas-Fermi approach and the kinetic energy is $$K = \frac{\hbar^2}{ 2m}  \left(\frac{\sigma}{d e}\right)^{2/3}$$ [see Fig. \ref{fig:phase_1}(a)].  On the other hand at   $\sigma \ll e/d^2$, the electrons are confined to the first sub-band of the triangular potential well, so that the  electron gas is  two dimensional (2DEG) \cite{RevModPhys.54.437}. The electron  kinetic energy is then $$K = \frac{\hbar^2}{2md^2}.$$ In both cases, the characteristic potential energy of electrons is $U = e E d$ [see Fig.\ref{fig:phase_1}(b)]. 

The dielectric response can be linear $E = D_0/\kappa$ or nonlinear $E = A D_0^3 /P_0^2 $ depending on whether the external electric field $D_0$ is smaller or larger than $D_c$, respectively. This gives us four cases which correspond to  high $\sigma \gg e/d^2$ and low $\sigma \ll e/d^2$ electron charge density, small field, $D_0 \ll D_c$, and large field, $D_0 \gg D_c$. The diagram in Fig. \ref{fig:phase_1} (c) shows the combined scaling results for $d(\sigma,D)$ for all these four domains. We use $\sigma$ as an independent coordinate because the three-dimensional concentration $\sigma/d$ can be obtained by  Hall effect measurements, while $d$ can also be measured independently  \cite{Hwang_PL}.

First, let us consider the domain of high electron charge density $\sigma \gg e/d^2$ and small field $D_0 \ll D_c$. The kinetic energy is $K = \hbar^2/2m (\sigma/e d)^{2/3}$ and the potential energy is $U = e  d D_0/\kappa$. From the condition $ K = U$ one obtains:
\begin{equation}
  \label{eq:small_high}
  d = b \kappa^{3/5} \frac{\sigma^{2/5}}{ D_0^{3/5}} \left(\frac{e}{b^2}\right)^{1/5}
\end{equation}
[see Fig. \ref{fig:phase_1}(c)]. When the neutrality relation (\ref{eq:neutrality}) is satisfied 
\begin{equation}
  \label{eq:critical_trivial}
\sigma = \frac{D_0}{4\pi} ,
\end{equation}
 we recover the  previous accumulation layer result, Eq. ~(\ref{eq:d_small}). For low electron charge density  $\sigma \ll e/d^2$ and small field $D_0 \ll D_c$ we get 
\begin{equation}
  \label{eq:small_low}
  d =  b \left(\frac{\kappa e/b^2 }{D_0}\right)^{1/3}.
\end{equation}
This is the classical result for the width of the inversion layer \cite{Stern_inversion}. The critical electron  density at which the transition from low to high electron charge density  occurs can be obtained from the condition $\sigma_c = e/d^2$ or from equating expressions (\ref{eq:small_high}) and (\ref{eq:small_low}) for $d$, giving 

\begin{equation}
  \label{eq:critical1}
  \sigma_c = \frac{e}{b^2} \frac{1}{\kappa^{2/3}} \left(\frac{D_0}{e/b^2}\right)^{2/3}.
\end{equation}

We emphasize the difference between the two regimes of low    [Eq. (\ref{eq:small_low})] and high [Eq. (\ref{eq:small_high})] electron charge densities. Sometimes~\cite{MacDonald_theory,induced_superconductivity},   Eq. (\ref{eq:small_low}) for $d$ of an inversion layer is used for an accumulation layer. 

So far we described the two left, low $D_0$ ($D_0<D_c = e/(a^2\sqrt{\kappa}) $) domains of Fig.~\ref{fig:phase_1}c.
For a large  field $D_0 \gg D_c$ and a  high electron charge density  $\sigma \gg e/d^2$  we get 
\begin{equation}
  \label{eq:large_high}
  d =  b  \left( \frac{b}{a} \right)^4 \frac{\sigma^{2/5}}{D_0^{9/5}} \left(\frac{e}{b^2}\right)^{7/5}.
\end{equation}
At $\sigma = D/4\pi $, i.e. when the neutrality relation is satisfied, we get the previous result Eq. (\ref{eq:d_STO}). 
At last, for a low electron charge density  $\sigma \ll e/d^2$ and large  field $D_0 \gg D_c$ we get 
\begin{equation}
  \label{eq:large_low}
  d = b \left( \frac{b}{a} \right)^{4/3} \frac{e/b^2}{D_0}.
\end{equation}
The transition from low to high electron charge density  at $D_0 \gg D_c$ occurs at the critical electron  density:
\begin{equation}
  \label{eq:critical2}
  \sigma_c = \frac{e}{b^2}  \left(\frac{D_0}{e/b^2}\right)^{2}.
\end{equation}
All four results Eqs. (\ref{eq:small_high}), (\ref{eq:small_low}), (\ref{eq:large_high}), (\ref{eq:large_low}) and all the border lines between different domains, Eqs. (\ref{eq:critical_trivial}), (\ref{eq:critical1}), (\ref{eq:critical2}), are shown  in Fig. \ref{fig:phase_1}c.

\section{Depletion layer}
Schottky diodes are metal-$n$-type-$\mathrm{SrTiO_3}$ junctions where the electron gas is depleted near the interface. Below we calculate the capacitance of this junction as a function of concentration of donors $N$ and applied voltage $V$, which has been experimentally studied \cite{Hwang_Schottky,Schottky_Hwang,Stemmer_Shottky,JJAP.36.L390}.

\label{sec:depletion}
\begin{figure}
\includegraphics[width=0.8\linewidth]{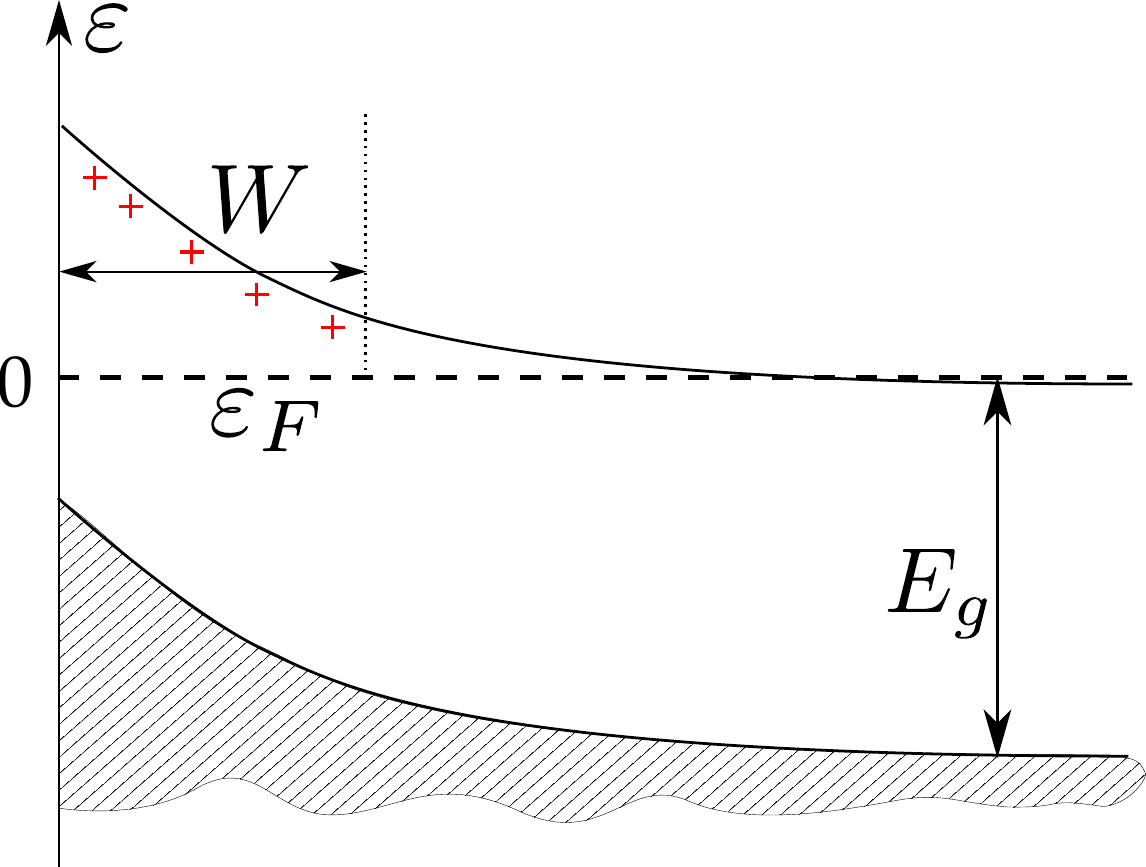}
\caption{(Color  online) Schematic energy diagram  of a depletion layer of $n$-doped semiconductor. Electrons have been forced away by an external electric field, which is created by negative charges  on the surface of the semiconductor. As a result, the depletion layer with width $W$   is filled by positively charged donors (red pluses).}
\label{fig:depletion}
\end{figure}

In order to calculate the capacitance of the Schottky diode let us consider Eqs. \eqref{eq:Gauss}. We use the full depletion approximation, in which  we assume that  $\mathrm{SrTiO_3}$ is fully depleted from electrons over a distance $W$ from the surface (Fig. \ref{fig:depletion}). The charge density in that region is due to the ionized donors with 3D concentration $N$:
\begin{eqnarray}
  \label{eq:D}
  \rho& =&  \frac{e}{a^3} (N a^3) ,  ~~ x<W,\\ \nonumber
 \rho &=& 0,~~ x>W.
\end{eqnarray} 
From equations \eqref{eq:D} and \eqref{eq:Gauss},  one can determine the dependence of the width $W$ on the potential  drop across the depletion layer $\Delta\varphi=\varphi(\infty) - \varphi(0) = V_0 - V > 0$, where the negative voltage $V$ is applied to the metal and $V_0$ is the potential difference between the work functions of $\mathrm{SrTiO_3}$ and the metal. If   $D_0 < D_c$ one can use the linear relation \eqref{eq:electric_field_small} to get:
\begin{equation}
  \label{eq:voltage_small}
  \Delta\varphi=  \frac{2\pi}{\kappa} \frac{e}{a} (Na^3) \left(\frac{W}{a}\right)^{2}.
\end{equation}

\begin{figure}[t]
\includegraphics[width=\linewidth]{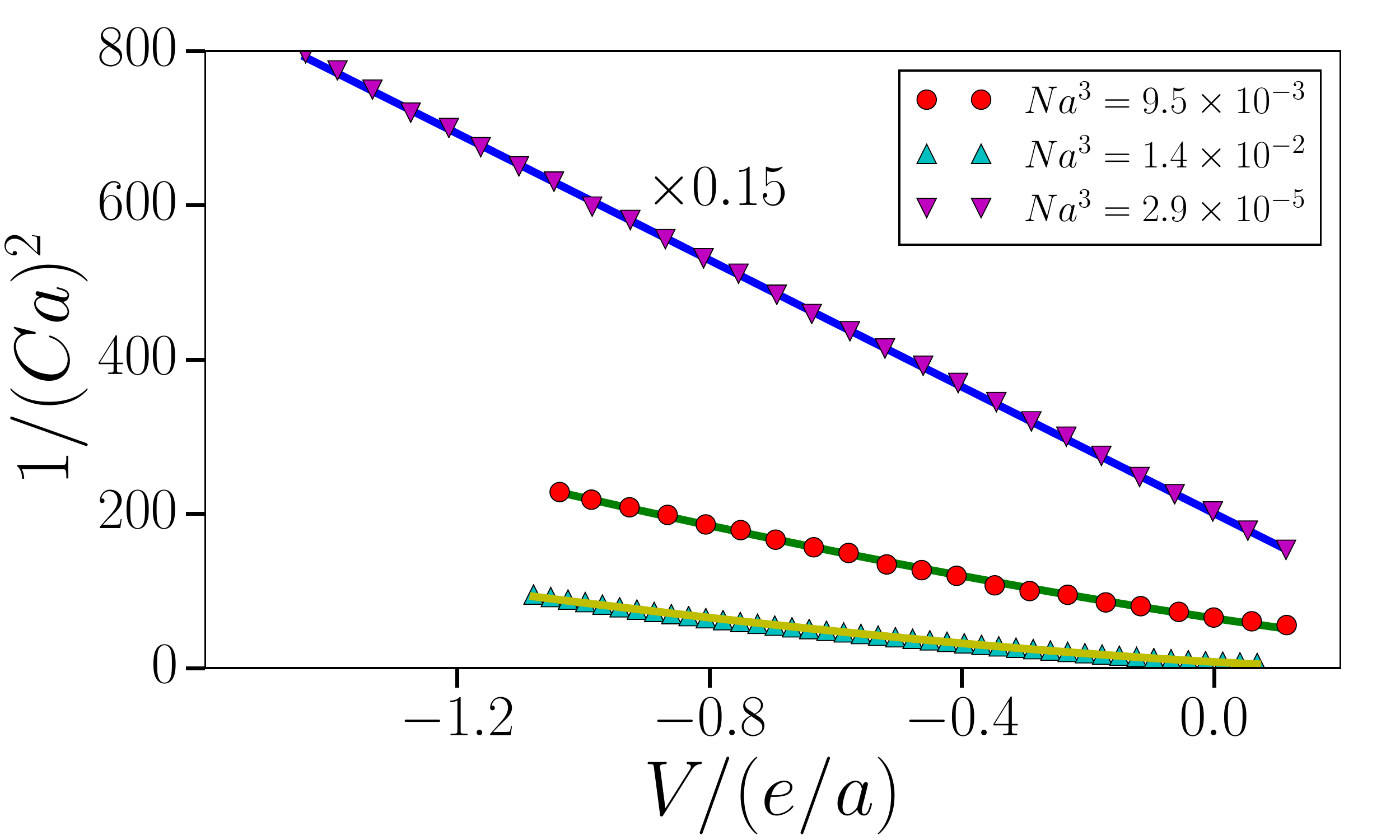}
\caption{(Color  online) Experimental capacitance-voltage (C-V) characteristics at room temperature  for  $\mathrm{Nb}$-doped $\mathrm{SrTiO_3}$ for three values of the donor concentration $N$.  The data for doping concentrations $Na^3=9.5\cdot 10^{-3}$ and $Na^3=2.9\cdot 10^{-5}$ are from Ref. \onlinecite{JJAP.36.L390}, the one for concentration $Na^3=1.4\cdot 10^{-2}$ are from Ref. \onlinecite{STO_Shottky_Mikheev} . The capacitance and the voltage are measured in dimensionless units $1/a$ and $e/a$, where $a$ is the lattice constant. Experimental results are shown by points. Solid lines correspond to fitting by Eq.~\eqref{eq:capacitance_small}  and by Eq. \eqref{eq:capacitance_high} for the smaller (upper line) and larger concentration (two lower lines), respectively.}
\label{fig:nonlinear_linear}
\end{figure}

If $D_0>D_c$, we use the nonlinear expression  \eqref{eq:electric_field_high}:
\begin{equation}
  \label{eq:voltage_high}
  \Delta\varphi= A \frac{1}{4} \frac{e}{a} (Na^3)^3 \left(\frac{W}{a}\right)^{4}.
\end{equation}
Now one can calculate the capacitance per unit area of the Schottky diode
$$
C=\frac{d Q}{ d \Delta\varphi} = e N \left |\frac{dW}{d \Delta\varphi}\right|. 
$$
This gives

\begin{equation}
  \label{eq:capacitance_small}
   \frac{1}{(Ca) ^2} = \frac{8\pi}{\kappa} \left(\frac{V_0-V}{e/a}\right)   \frac{1}{Na^3} 
\end{equation}
for $D_0<D_c$, and 
\begin{equation}
  \label{eq:capacitance_high}
   \frac{1}{(Ca) ^2} = 8 \left(\frac{V_0-V}{e/a}\right)^{\frac{3}{2}} \left( \frac{A}{Na^3} \right)^{\frac{1}{2}}
\end{equation}
for $D_0>D_c$. With growing $V$ the cross-over between Eqs. (\ref{eq:capacitance_small}) and (\ref{eq:capacitance_high}) happens at $V=V_c$ where:
\begin{equation}
  \label{eq:Vc}
  V_c=V_0-\frac{e}{a}\left( \frac{\pi}{\kappa}\right)^2 \frac{1}{A Na^3}.
\end{equation}


To test our theoretical predictions we consider the experimental data  obtained at room temperature  for lightly and heavier  $\mathrm{Nb}$-doped  $\mathrm{SrTiO_3}$ of Ref. \cite{JJAP.36.L390}, $Na^3=2.9\cdot 10^{-5}$  and $Na^3=9.5\cdot 10^{-3}$  as well as the more heavily doped sample  $Na^3=1.4\cdot 10^{-2}$ of Ref. \onlinecite{STO_Shottky_Mikheev} (see Fig. \ref{fig:nonlinear_linear}). In the first case, one can expect $D_0$ to be small and we  use  Eq.~\eqref{eq:capacitance_small} to fit the experimental data. Using  $\kappa$ and $V_0$ as fitting parameters we get $V_0=0.5 e/a$, $\kappa \simeq 320$, which is close to the room temperature value $\kappa \simeq 350$ for $\mathrm{SrTiO_3}$. For the heavier doped  samples, one can expect that  $D_0$ is large and  use Eq. \eqref{eq:capacitance_high} to find $A$ and $V_0$ from the  experimental data. As a result, we get $V_0=0.8 e/a$, $A \simeq 1.5$ and $V_0=0.26 e/a$, $A \simeq 0.8$ for the data of  Refs. \onlinecite{JJAP.36.L390} and \onlinecite{STO_Shottky_Mikheev}, respectively.  \footnote{One can verify whether  Eqs. (\ref{eq:capacitance_small}) and (\ref{eq:capacitance_high}) are applicable for describing the capacitance-voltage characteristics.  To do this we consider the critical voltage (\ref{eq:Vc}). For small concentration $V_c \simeq -2 e/a$ is smaller than the experimental voltage $V$ (see Fig.  \ref{fig:nonlinear_linear}). It means that using  Eq. (\ref{eq:capacitance_small}) is justified. For high concentrations $V_c \simeq 0.8 e/a$ or $0.26 e/a$ is larger than the experimental voltage $V$. For such voltages one can use Eq. (\ref{eq:capacitance_high}).}

In Refs.  \onlinecite{Berg_Schottky_STO,JJAP.36.L390,Shotky_Landau_starnge_formula,STO_Shottky_Mikheev,Kahng_polarization} $C(V)$ data are used  to derive $\kappa(E) \equiv d D/d E$, which agrees with the Landau description \cite{Shotky_Landau_starnge_formula}. Extracting the parameter $A$ from $\kappa(E)$ of Ref. \onlinecite{Berg_Schottky_STO} gives  $A=0.5$, while $\kappa(E)$ in Refs. \onlinecite{JJAP.36.L390,STO_Shottky_Mikheev} lead to the above mentioned values of $A$. The scatter of values of $A$ probably can be explained by  the effect of a non-controllable ``dead layer'' of low dielectric constant between $\mathrm{SrTiO_3}$ and the metal \cite{JJAP.36.L390,STO_Shottky_Mikheev}. 

It is believed \cite{STO_Shottky_Mikheev} the best interface was made in Ref. \onlinecite{STO_Shottky_Mikheev}, whose data lead to $A \simeq 0.8$.  Above we used this value for all numerical estimates. \footnote{Note that there are several other methods \cite{Neville_Permitivity_STO,Electric-Field-Induced_Raman_Scattering} to study the non-linear dielectric response, but they deal with relatively weak electric fields, while we are interested in strong electric fields.}

\section{Does spatial dispersion of dielectric response  affect the accumulation layer?}
\label{sec:dispersion}
So far we assumed that the external field $D_0$ and polarization $P$ change abruptly  at the interface, i.e., we have ignored the dispersion of the dielectric response.  In this section we show that even without this assumption, the results of previous sections remain intact for $\mathrm{SrTiO_3}$. We concentrate on an accumulation layer which can be so narrow that the question of spatial dispersion arises. To take into account that the electric field and polarization can not change abruptly, in  the geometry chosen  above we add the gradient term $(1/2) a_0^2(dP/dx)^2$ to the Landau-Ginzburg free energy density Eq. (\ref{eq:F}). Here $a_0^2=g_{xxxx}$ in the general expression for the gradient term
$$
\frac{1}{2}g_{iklm}\frac{\partial P_i}{\partial x_k} \frac{\partial P_l}{\partial x_m},
$$
 and  $i,k,l,m$ enumerate the three coordinates $x,y,z$. Adding such a term necessitates an additional boundary condition for $P(x)$ at $x=0$. The general form of this condition~\cite{Kretschmer_boundary} is $P + \lambda dP/dx =0$. It does not bring new physics into the problem if $\lambda$ is large. Therefore, we explore the opposite case where $\lambda$ can be ignored, $P(0)=0$. From the condition $\partial F/\partial P =0$  we get:
\begin{equation}
 \label{eq:electric_field_dispersion}
 E = \frac{4\pi}{\kappa} P + \frac{A}{P_0^2} P^3 - a_0^2 \frac{d^2 P}{d x^2}. 
\end{equation}
Let us first find a solution to Eqs. (\ref{eq:Gauss}) together with Eq. (\ref{eq:electric_field_dispersion})  for a  system  without electrons ($\rho(x)=0$). The electrical induction $D_0= E+ 4\pi P$ is constant everywhere 
 and we get an approximate solution for  $E=D_0-4\pi P$:
\begin{equation}
  \label{eq:electric_field_dispersion_answer}
E \simeq (D_0-E_{\infty}) \exp\left( - \sqrt{4\pi} \frac{x}{a_0} \right) + E_\infty,
\end{equation}
where  $E_\infty=D_0/\kappa + (D_0/4\pi)^3/P_0^2$ is the electric field in the bulk. 

Now, we take electrons into account and start from  the  case where in Eq. (\ref{eq:electric_field_dispersion})  $\kappa=\infty$ and $P_0=\infty$, so that the role of the gradient term is emphasized. From the resulting  equation $E=-a_0^2d^2P/dx^2$ we get $\varphi=a_0^2 dP/dx$. (We take into account that at $ x\rightarrow \infty $, $\varphi = dP/dx=0 $.) From the condition $dE/dx+4\pi dP/dx = 4\pi e \rho$ we get an equation for the potential:
\begin{equation}
  \label{eq:potential_dispersion}
\frac{d^2 \varphi}{dx^2} = \frac{4\pi}{a_0^2}\varphi + \frac{e}{b^3} \frac{2^{7/2}}{3\pi} \left(\frac{\varphi}{e/b}\right)^{3/2}
\end{equation}
(compare with Eq.~(\ref{eq:potential_equation_linear})). The solution to this equation, with boundary condition  $\varphi(\infty)=0$  is:

\begin{equation}
  \label{eq:potential_dispersion}
\varphi(x) = \frac{15^2\pi^4}{2^{7}} \frac{e b^3}{a_0^4}\frac{1}{\sinh^4 X},
\end{equation}
where  $X=\sqrt{\pi}(x+d)/2 a_0$ and $d$ is still an unknown length, which can be found from the boundary condition. Let us describe this solution. First, we see that the solution critically depends on the parameter $X$. If $X>1$ or $x>a_0 - d$ then one arrives at Eq. (\ref{eq:electric_field_dispersion_answer}): the electric field and potential decay exponentially to zero as $\propto \exp(- \sqrt{4\pi} x/a_0)$.

Let us consider the case when $X<1$ or $x<a_0 - d$. In that case we get a result similar to  Eq. (\ref{eq:potential_linear}), but with $\kappa=1$
\begin{equation}
  \label{eq:small_alpha}
\varphi=\frac{225 \pi^2}{8} \frac{e b^3}{(x+d)^4}.  
\end{equation}
Thus, at  $x<  a_0-d$ the electrons screen the external field much faster than the lattice, and hence the latter does not participate appreciably in the screening response.


One can try to estimate  $d$ assuming  that  the majority of  electrons are in the region  $0<x<a_0-d$. In that case the potential is given by Eq. (\ref{eq:small_alpha}) and  $d$ can be found from the condition of neutrality (\ref{eq:neutrality}):

\begin{equation}
  \label{eq:d_small_without_kappa}
  d_1= C_5 b   \left(\frac{e/b^2}{D_0}\right)^{1/5}.
\end{equation}

The condition $d_1<a_0$ required for applicability of Eq. (\ref{eq:small_alpha}), is valid only when $D_0 \gg D_{d}$, where 
\begin{equation}
  \label{eq:D_critical_dispersion}
  D_d=C_5^5 b^3 \frac{e}{a_0^5} \simeq 1000 \left(\frac{a}{a_0}\right)^5 \left(\frac{b}{a}\right)^3 \frac{e}{a^2}.
\end{equation}

Let us now consider  the opposite case  $D_0 \ll D_d$, where most of the electrons are not located  in the region $x<a_0-d$. In fact, to find where they are located one has to introduce finite $\kappa$ and $P_0$. The exponential decay of the electric field with $x$ saturates at the level of $E_\infty$. At larger $x$ we reach the electron gas with a larger width given by Eqs. ~(\ref{eq:d_small}) or (\ref{eq:d_STO}) for linear and nonlinear dielectric responses respectively. In this case, the electron gas resides at large distances and our considerations of the previous section are valid.

If we return to  finite $\kappa$ and $P_0$  in Eq. (\ref{eq:electric_field_dispersion})  in the first  case $D_0 \gg D_d$,  we arrive to following hybrid picture.  An electron gas screens most of the external field at small $x\simeq d_1$. At larger distances we again get exponential decay described by Eq.~(\ref{eq:electric_field_dispersion_answer}), which results from the lattice response. After  this exponential decay the electron concentration $n(x)$ again follows Eq. (\ref{eq:conventration_linear})  or Eq. (\ref{eq:conventration_nonlinear}), with the bulk value of the dielectric constant  $\kappa$. In other words, the electron density has a two component distribution. The electrons nearest to the interface are likely localized due to disorder. This could, in principle, create hope to explain why at the $\mathrm{LaAlO_3/SrTiO_3}$ interface  only 10\% of the electrons predicted by the polar catastrophe, $\sigma=0.5. e/a^2$, are observed in transport measurements. 

However, this hope does not survive for actual parameters of $\mathrm{SrTiO_3}$, $b=0.35 \mathrm{\AA},$  $a_0 = 1 \mathrm{\AA}$ \cite{Tagantsev_background_dielectric_constant} . Although for such parameters the Thomas-Fermi approximation is still valid, the electric field at which all electrons are located within the distance $a_0$ from the interface, $D_d = 650 e/a^2$ is so large  that $D_0 \ll D_d$  always. One may also worry that for such values of  $b$ and $a_0$, which are smaller than the lattice constant $a$ of $\mathrm{SrTiO_3}$, the continuous theory that we use is not applicable.
 
However, the applicability of the continuous theory can be much better due to  the so called background dielectric constant~\cite{Tagantsev_background_dielectric_constant}. Indeed, until now we have described the entire dielectric response by the Landau-Ginzburg theory for a single order parameter, which can be identified with the displacement of  the  transversal optical soft mode of $\mathrm{SrTiO_3}.$ This is a good approach when the dielectric response is very strong. However, near the interface the response of the soft mode is weakened due to the dispersion and the response of other optical modes as well as the polarization of ions must be included. This is done~\cite{Tagantsev_background_dielectric_constant}  by the addition of the linear non-dispersive background dielectric constant, $10<\kappa_b< 30$~\cite{Tagantsev_background_dielectric_constant,Shannon_dielectric_pervovskite}. To model this situation one can  replace  $P$ by the soft mode contribution $P_s$ in the free energy density (\ref{eq:F}) and add to it $(2\pi /\kappa_b)P_b^2 -E P_b$ while keeping  $P = P_s + P_b$ in Eqs.  (\ref{eq:Gauss}). Here $P_b$ is the background polarization.

As a result, the small distance dielectric constant becomes $\kappa_b$ instead of 1 and the lengths $b$, $a_0$ and the characteristic field of Eq. (\ref{eq:D_critical_dispersion}) $D_d$ are replaced by $b\kappa_b$, $a_0 \sqrt{\kappa_b}$ and $D_d \kappa_b^{1/2}$, respectively. For $10<\kappa_b< 30$ the characteristic lengths $b\kappa_b$ and  $a_0 \sqrt{\kappa_b}$ may  reach the lattice constant $a$, thereby improving the applicability of our theory. At the same time the critical field $D_d$  becomes even larger so that in all realistic situations, $D_0 \ll D_d$,  all electrons are located at distances larger than  the lattice constant $a$. This means that the dispersion does not play a substantial role and  the dispersion-less approach used in the previous sections is applicable for describing accumulation, inversion and depletion layers in $\mathrm{SrTiO_3}$.

\section{Conclusion}
\label{sec:conclusion}

In this paper we studied the potential and electron density depth profiles  in accumulation,  inversion and depletion layers for materials  with a very large dielectric constant and nonlinear dielectric response such as $\mathrm{SrTiO_3}$. In particular, we showed that in a depletion layer at a given donor concentration and for high enough voltage, the dependence of the capacitance on voltage decreases as $C \propto V^{-3/4}$, which is substantially different from the conventional result $C \propto V^{-1/2}$ with linear dielectric response. For an inversion layer we found that the layer width depends on the external electric field $D_0$ as  $d \propto (\kappa/D_0)^{1/3}$ and $d \propto 1/D_0$ for linear and nonlinear dielectric responses, respectively. In accumulation layers near interfaces like $\mathrm{GdTiO_3/SrTiO_3}$, $\mathrm{LaTiO_3/SrTiO_3}$, and $\mathrm{LaAlO_3/SrTiO_3}$ we obtained $n(x) \propto (x+d)^{-12/7}$ with $d \propto D_0^{-7/5}$, due to the nonlinearity of the dielectric response. We found that 70\% of electrons are located within $2.6~\mathrm{nm}$ of the interfaces $\mathrm{GdTiO_3/SrTiO_3}$ and $\mathrm{LaTiO_3/SrTiO_3}$ (where the electron surface charge density is $\sigma = 0.5 e/a^2$) in  agreement with experimental data. The predicted functional shape of the electron depth profile $n(x)$   also shows satisfactory agreement with the experimental data. Spatial dispersion in the dielectric response was shown to be negligible for the description of potential and electron density depth profiles  in $\mathrm{SrTiO_3}$ devices. This paper uses a simplified isotropic electron spectrum, while the electronic structure of $\mathrm{SrTiO_3}$ is multiorbital in nature. In future work we plan to go beyond our simplified description.

\begin{acknowledgments}
We are grateful to A. P. Levanyuk and S. Stemmer for reading the manuscript and attracting our attention to a number of important references. We thank R. M. Fernandes, B. Jalan, A. Kamenev, B. Skinner, C. Leighton, A. J. Millis, A. Sirenko and P. W{\"o}lfle for helpful discussions. This work was supported primarily by the National Science Foundation through the University of Minnesota MRSEC under Award Number DMR-1420013.
\end{acknowledgments}

\end{document}